\documentclass[pdflatex,sn-mathphys-num]{sn-jnl}% M

\usepackage{graphicx}%
\usepackage{multirow}%
\usepackage{amsmath,amssymb,amsfonts}%
\usepackage{amsthm}%
\usepackage{mathrsfs}%
\usepackage[title]{appendix}%
\usepackage{xcolor}%
\usepackage{textcomp}%520
\usepackage{manyfoot}%
\usepackage{booktabs}%
\usepackage{algorithm}%
\usepackage{algorithmicx}%
\usepackage{algpseudocode}%
\usepackage{listings}%
\usepackage{tikz}
\definecolor{lime}{HTML}{A6CE39}
\DeclareRobustCommand{\orcidicon}{
	\begin{tikzpicture}
		\draw[lime, fill=lime] (0,0) 
		circle [radius=0.16] 
		node[white] {{\fontfamily{qag}\selectfont \tiny ID}};
		\draw[white, fill=white] (-0.0625,0.095) 
		circle [radius=0.007];
	\end{tikzpicture}
	\hspace{-2mm}
}

\newcommand{\orcidlink}[1]{\href{https://orcid.org/#1}{\orcidicon}}

\unnumbered% uncomment this for unnumbered level heads

\begin{document}

\title[Article Title]{Single-Shot Lensless Imaging with Physics Guided Genetic Programming}

\author[1]{\fnm{Ganesh M.} \sur{Balasubramaniam\orcidlink{0000-0001-7824-5831}}}
\author[2]{\fnm{Xiao-Liu} \sur{Chu\orcidlink{0000-0002-7476-2097}}}
\author[2]{\fnm{Radhika V.} \sur{Nair\orcidlink{0000-0002-4370-2112}}}
\author*[1,2]{\fnm{Matthew R.} \sur{Foreman\orcidlink{0000-0001-5864-9636}}}\email{matthew.foreman@ntu.edu.sg}

\affil[1]{School of Electrical and Electronic Engineering, Nanyang Technological University, 50 Nanyang Avenue, Singapore 639798}

\affil[2]{Institute for Digital Molecular Analytics and Science, Nanyang Technological University, 59 Nanyang Drive, Singapore 636921}

\abstract{Lensless optical imaging eliminates the need for refractive optics, enabling compact and low-cost cameras with a large field-of-view, supporting point-of-care diagnostics and industrial monitoring. Practical deployments, however, remain constrained by ill-posed image reconstruction pipelines that require multiple measurements, careful calibration or object-specific training, thus limiting robustness and scalability. In this work, we introduce a single-shot lensless imaging framework that reconstructs complex objects from only a single recorded intensity pattern using a genetically programmed iterative algorithm. Our method couples a wave-propagation model with an adaptive meta-optimisation strategy to jointly estimate the object amplitude, object phase, and effective object-detector distance. Experiments demonstrate high-fidelity recovery of amplitude objects, including a USAF target and 2~\(\mu\)m silicon beads on a glass slide, as well as a phase-dominant biological sample consisting of U2OS cells on a glass slide. Across multiple object types, wavelengths, and propagation distances, the same learned policy maintains high reconstruction quality with minimal retuning, indicating strong out-of-distribution generalisation. As a practical demonstration, the framework is integrated with a $\beta$-amyloid-based optical digital bead assay under wide field-of-view acquisition. The resulting platform combines single-shot capture, compact hardware, and accurate reconstruction of complex fields, enabling rapid, portable assays in which throughput, alignment tolerance, and cost are critical.}


\maketitle

\section{Introduction}\label{intro}

Single-shot lensless optical imaging aims to reconstruct an object from a single intensity measurement recorded directly on an image sensor \cite{Boominathan2022OpticaReview}. This architecture offers great promise for field-deployable high-throughput biomedical imaging systems, including digital bead assays, cytometers, and cell counting assays, because it enables wide-field acquisition with compact hardware and minimal alignment overhead \cite{Ozcan2016Lensless,rosen2024roadmap,potter2024clinical,Seo2009Lensfree,Im2015D3}. Single-shot image recovery, however, is fundamentally ill-posed. Practical deployments must contend with twin image artefacts in inline holography, sensitivity to object-sensor separation, reduced effective resolution under pixel-limited sampling, and measurement variability driven by noise, contrast, and sample-dependent scattering \cite{Shechtman2015PR,Arcab2024TwinImage}. These limitations are especially detrimental in assay readouts, where systematic reconstruction bias can translate into counting errors or false positives. 

Physics-driven reconstruction methods can be used to address the issue of the ill-posed inverse problem by enforcing wave optics consistency through iterative phase retrieval or by introducing controlled measurement diversity \cite{Gerchberg1972GS,Fienup1982PhaseRetrieval,Shechtman2015PR}. While schemes to enhance data diversity, such as multi-height, multi-plane or ptychographic acquisition, can improve robustness, they often do so at the cost of increased acquisition complexity and reduced throughput \cite{Greenbaum2012MH,Maiden2009PIE,Zheng2013FPM}. Alternatively, coded aperture and diffuser-based architectures, such as FlatCam and DiffuserCam, enable single-shot capture in thin form factors, yet these methods typically rely on calibration and carefully tuned inverse solvers, which can degrade under drift or sample-induced mismatch \cite{Asif2017FlatCam,Antipa2018DiffuserCam}. To further stabilise reconstructions, regularised optimisation using priors such as total variation and sparsity is often employed. However, performance remains sensitive to hyperparameter choices and stopping criteria, particularly when defocus and noise vary across measurements \cite{Rudin1992ROF,Rivenson2016MH}. These constraints have recently spurred increased interest in learning-based methodologies that minimise manual adjustment while maintaining reconstruction efficiency across imaging settings.

Learning-based reconstruction offers an alternative approach to single-shot image recovery by learning implicit priors from data \cite{Rivenson2018LSA,Wu2018HIDEF,Wang2018eHoloNet,Li2020DeepDIH}. In this context, learned models are attractive because they can absorb part of the reconstruction burden that would otherwise be handled through explicit regularisation design and solver configuration. Untrained and physics-guided networks reduce dependence on paired datasets by optimising network parameters against a forward model, while diffusion-based priors have recently been coupled to holographic models to improve robustness \cite{Bostan2020Optica,Wang2020PhysenNet,Zhang2024PadDH,Mandal2023SciRep}. Nonetheless, learning-based methods can exhibit sensitivity to distribution shifts across sample classes and optical configurations, and stable performance often requires careful selection of regularisation weights and update schedules \cite{Rivenson2018LSA,Bostan2020Optica}. Recent physics-driven neural formulations have also highlighted the continuing uncertainty between single-shot capability and robustness to experimental variability \cite{Kim2025MorpHoloNet}.

The limitations discussed above point to a persistent requirement for an image reconstruction strategy that remains physically grounded, avoids multiple measurements and large object-specific training datasets, and performs robustly across varying samples and acquisition conditions. This study presents a physics-informed genetic programming (GP) framework that configures an iterative wave-optics-based inverse solver for single-shot lensless image reconstruction \cite{Koza1992GP,Poli2008FieldGuideGP,EibenSmith2015EC,Khan2021GPSurvey}. The main innovation in the proposed method is that the GP is not used to learn a direct image-to-image reconstruction model. Rather, it adapts the optimization process to the measurement by evolving an interpretable symbolic policy that maps low-dimensional descriptors of the recorded intensity to the bounded inverse solver hyperparameters. In this way, the framework retains a fully physics-based reconstruction pipeline while reducing the manual hyperparameter tuning that has historically limited the robustness and transferability of iterative lensless solvers. The evolved policy governs learning rate schedules, regularisation weights, and stopping criteria, and is trained using a standard USAF resolution target. It is then evaluated on distinct sample classes, including intensity measurements of 2~\(\mu\)m silicon microparticles and human osteosarcoma cells, thereby probing generalisation beyond the training target. Across these regimes, the evolved configuration improves convergence stability and reconstruction fidelity, supporting reliable recovery from a single measurement under practical variations in noise, contrast, and defocus. The same reconstruction strategy is subsequently applied to a $\beta$-amyloid-based digital bead assay under wide field-of-view acquisition, enabling rapid, portable and high field-of-view readout of particles in settings where throughput, alignment tolerance, and cost are critical \cite{Boominathan2022OpticaReview,Ozcan2016Lensless,potter2024clinical}.

\section{Results}
\label{secResults}

\subsection{Lensless Imaging Measurements}
\label{ImforLen}

\begin{figure}[t]
    \centering
    \includegraphics[width=\linewidth]{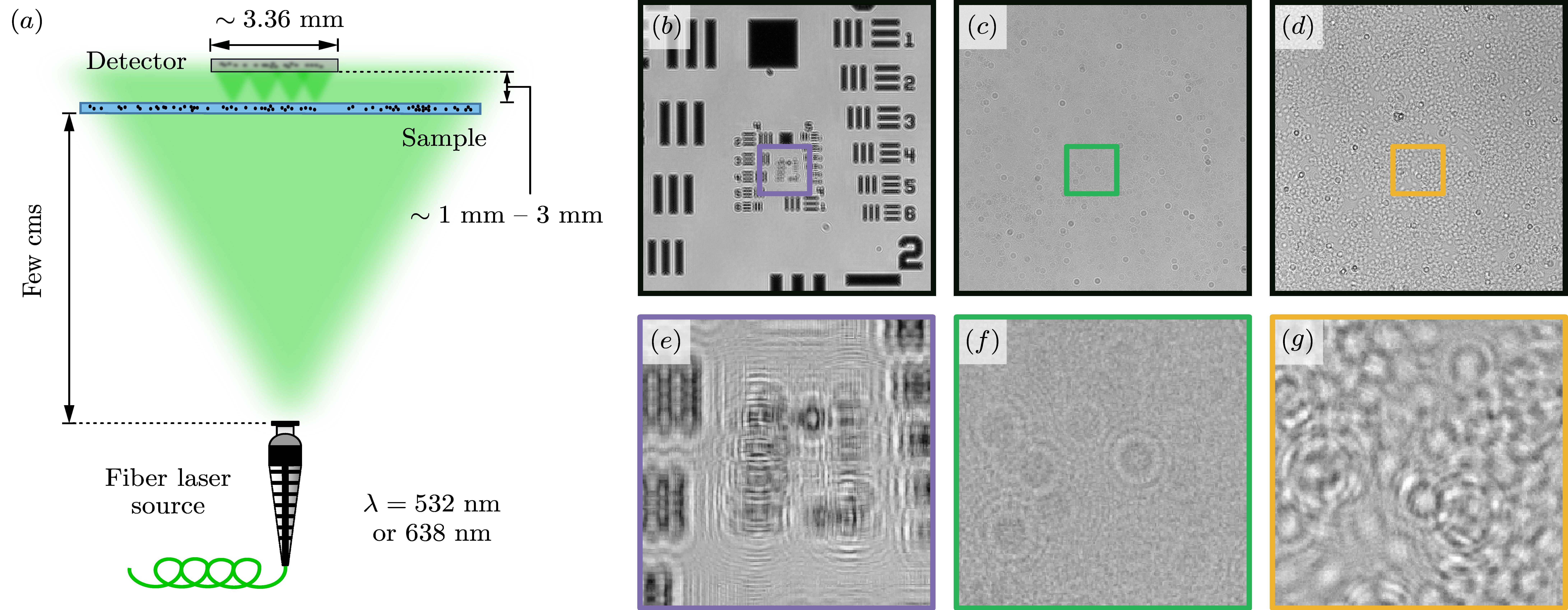}
    \caption{\textbf{Lensless imaging configuration and representative single-shot measurements.} (a) An inline lensless setup in which the sample is illuminated by a monochromatic laser source and a 3.36~mm $\times$ 3.36~mm sensor records the diffracted intensity without imaging optics. Single-shot intensity measurement of: (b) a USAF resolution target measured using a 520~nm source; (c) randomly dispersed 2~\(\mu\)m silicon microparticles measured using a 520~nm source; and (d) human osteosarcoma (U2OS) cells measured using a 638~nm source. (e)--(g) Enlarged images corresponding to regions shown in (b)-(d) respectively showing diffraction fringes and self-interference structure that preclude direct interpretation of the sample from raw intensity data. }
    \label{fig:schematic_meas}
\end{figure}

The lensless imaging configuration used in this study is shown in Fig.~\ref{fig:schematic_meas}(a) and is described in the Methods Section. The system adopts a compact inline geometry in which a spatially extended, monochromatic laser illuminates the sample, and the sensor records the resulting intensity distribution directly, without the use of additional refractive optics. The sample is placed at a finite propagation distance from the detector, so the recorded measurement is a Fresnel diffraction pattern rather than an in-focus image. The propagation distance is not assumed to be known with sufficient accuracy for reconstruction and is treated as an unknown to be estimated by the computational pipeline.

The recorded intensity encodes object information through interference between a background unscattered reference field and the field scattered by the object, which produces diffraction fringes and self-interference features \cite{Ozcan2016Lensless,Boominathan2022OpticaReview,Shechtman2015PR}. In an inline configuration, these effects can lead to severe ambiguities in single-shot acquisition, including strong sensitivity to defocus. Furthermore, the absence of optical magnification couples sensor sampling, object size, and propagation distance, which can suppress high spatial frequencies and reduce effective resolution when reconstruction is not carefully regularised \cite{Shechtman2015PR,rosen2024roadmap}.

Representative single-shot measurements of both amplitude and phase objects acquired in this study are shown in Fig.~\ref{fig:schematic_meas}(b) to Fig.~\ref{fig:schematic_meas}(d), with enlarged regions in Figs.~\ref{fig:schematic_meas}(e), ~\ref{fig:schematic_meas}(f), and ~\ref{fig:schematic_meas}(g) highlighting typical degradations. The amplitude-dominated measurements, performed using a 520~nm source, include a USAF resolution target in Fig.~\ref{fig:schematic_meas}(b) and a microparticle sample in Fig.~\ref{fig:schematic_meas}(c). The phase-dominated biological measurement, performed using a 638~nm source, is shown in Fig.~\ref{fig:schematic_meas}(d). Across these cases, the raw intensity patterns are dominated by diffraction fringes and defocus-dependent structure, which renders direct visual interpretation infeasible.

The described measurements are intentionally diverse, spanning amplitude-dominant and phase-dominant objects as well as distinct wavelengths and object-detector propagation distances. This diversity provides a stringent basis for evaluating whether a single reconstruction strategy remains effective across different contrast mechanisms and sample statistics under single-shot acquisition. It also directly reflects the practical operating conditions under which lensless systems must accommodate heterogeneous specimens and modest variability in object-sensor spacing without requiring repeated recalibration. 

\subsection{Image Reconstruction Procedure and Cross-Sample Evaluation}
\label{ImreconAll}

The reconstruction task is to recover the object field, or more specifically its amplitude \(a(x_{\mathrm{o}},y_{\mathrm{o}})\) and phase \(\phi(x_{\mathrm{o}},y_{\mathrm{o}})\), together with an effective object-sensor propagation distance \(z\), from a single measured intensity image \(I(x_{\mathrm{s}},y_{\mathrm{s}})\). With a suitably well-characterised illumination field, the recovered complex object field can in principle also be related to the object transmission coefficient and, for appropriate samples, to optical thickness. In the present study, however, the focus is on robust single-shot reconstruction and assay-relevant readout. The inverse problem is posed through a wave optics forward model
\begin{equation}
I(x_{\mathrm{s}},y_{\mathrm{s}}) =
\left|\mathcal{P}_{z}\left\{a(x_{\mathrm{o}},y_{\mathrm{o}})\exp\left(i\phi(x_{\mathrm{o}},y_{\mathrm{o}})\right)\right\}(x_{\mathrm{s}},y_{\mathrm{s}})\right|^{2},
\label{eq:forward_model}
\end{equation}
where \(\mathcal{P}_{z}\{\cdot\}\) denotes free space propagation over distance \(z\) from the object plane coordinates \((x_{\mathrm{o}},y_{\mathrm{o}})\) to the sensor plane coordinates \((x_{\mathrm{s}},y_{\mathrm{s}})\) and the goal is to achieve reliable single-shot recovery across distinct object classes and wavelengths without per-sample recalibration or manual retuning.

Image reconstruction is performed using the physics-informed evolutionary framework described in the Methods Section. The algorithm operates on a single recorded intensity measurement and does not require ground-truth images or supervised training data. A measured image of size \(2400 \times 2400\) pixels is partitioned into \(4 \times 4\) tiles, each with a size of \(600 \times 600\) pixels. With sensor pitch \(d_x = 1.4~\mu\mathrm{m}\), the field-of-view is \(N d_x \approx 3.36~\mathrm{mm}\) per side. All detector plane comparisons are computed using intensities normalised to \([0,1]\). During optimisation, the object field amplitude is constrained such that \(a_{ij}\in[0,1]\). Phase is treated as an unconstrained variable and is shown after linear normalisation to \([0,1]\) for visualisation where indicated.

\begin{figure}[p]
    \centering
    \includegraphics[width=\linewidth]{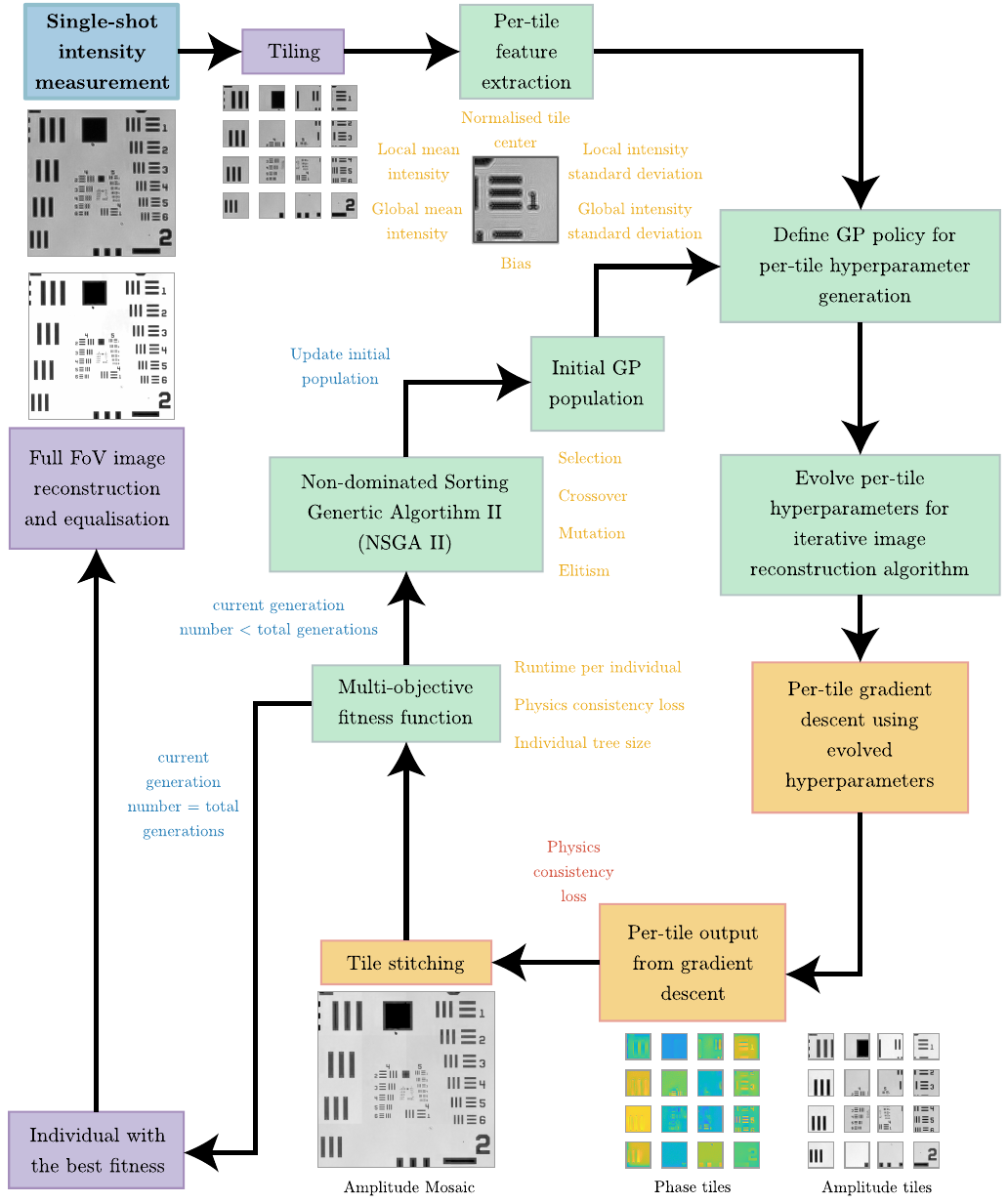}
    \caption{\textbf{Algorithmic flow of the proposed single-shot lensless reconstruction framework.} A single recorded intensity measurement is partitioned into tiles. Summary features are extracted per tile and mapped by a genetic programming policy to bounded hyperparameters. These hyperparameters configure a physics-based iterative solver that reconstructs tile-wise amplitude, phase, and propagation distance. Tile reconstructions are stitched to produce full-field estimates. Details of the outer policy evolution (green boxes) and inner reconstruction procedure (yellow boxes) are provided in the Methods.}
    \label{fig:GP_BlockDiag}
\end{figure}

The method proposed here has two associated levels. At the outer level, genetic programming evolves a policy that maps summary features of each measured tile to the hyperparameters of an inner physics-based solver. For tile \((i,j)\), the feature vector is
\begin{align}
\mathbf{f}_{ij} =
\bigl(X_{ij}, Y_{ij}, m_{ij}, s_{ij}, g_M, g_S, C\bigr),
\label{eq:tile_features}
\end{align}
where \((X_{ij},Y_{ij})\) are the normalised coordinates of the tile center, \(m_{ij}\) and \(s_{ij}\) are the mean and standard deviation of the measured tile intensity, whereas \(g_M\) and \(g_S\) are the corresponding global statistics of the full image. The constant \(C\) is a fixed bias input, common to all tiles, that allows constant offsets in the policy mapping. A policy \(\pi_{\boldsymbol{\psi}}\) with parameters \(\boldsymbol{\psi}\) outputs a bounded hyperparameter set
\begin{equation}
\boldsymbol{\theta}_{ij} = \pi_{\boldsymbol{\psi}}\left(\mathbf{f}_{ij}\right),
\label{eq:policy_map}
\end{equation}
which includes the iteration budget, learning rates, regularisation weights, the Huber parameter, and an initial guess of the object-detector propagation distance (see Methods for details). Policy evolution is driven by physics-based reconstruction quality and computational cost. Candidate policies are represented as sets of symbolic expression trees, and selection and variation are applied to these policies as described in the Methods Section.

At the inner level, tile-wise optimisation estimates a spatially dependent complex object field and an effective object-sensor propagation distance. The object field is parameterised as
\begin{align}
\tau_{ij}(x_{\mathrm{o}},y_{\mathrm{o}}) =
a_{ij}(x_{\mathrm{o}},y_{\mathrm{o}})
\exp\bigl(i\phi_{ij}(x_{\mathrm{o}},y_{\mathrm{o}})\bigr),
\label{eq:object_param}
\end{align}
where \(a_{ij}(x_{\mathrm{o}},y_{\mathrm{o}})\in[0,1]\) is the amplitude and \(\phi_{ij}(x_{\mathrm{o}},y_{\mathrm{o}})\) phase. The predicted detector plane intensity is obtained by a wave optics forward model, specifically
\begin{equation}
\hat{I}_{ij}(x_{\mathrm{s}},y_{\mathrm{s}})
=
\left|
\mathcal{P}_{z_{ij}}\left\{\tau_{ij}(x_{\mathrm{o}},y_{\mathrm{o}})\right\}(x_{\mathrm{s}},y_{\mathrm{s}})
\right|^{2}.
\label{eq:pred_intensity}
\end{equation}
The propagation distance, $z_{ij}$, is treated as unknown and constrained to a physically plausible interval using a sigmoid reparameterisation,
\begin{equation}
z_{ij} = z_{\min} + (z_{\max}-z_{\min})\, \sigma\left(\zeta_{ij}\right),
\label{eq:z_reparam}
\end{equation}
where \(\zeta_{ij}\) is an unconstrained optimisation variable and \(\sigma(\cdot)\) is the logistic function. In the reconstructions reported here, \(z_{\min}=1.34~\mathrm{mm}\) and \(z_{\max}=1.75~\mathrm{mm}\), which is consistent with the independently measured object-detector separations in the corresponding experimental configurations.

Given a measured tile intensity \(I_{ij}(x_{\mathrm{s}},y_{\mathrm{s}})\) normalised to \([0,1]\) and a predicted intensity \(\hat{I}_{ij}(x_{\mathrm{s}},y_{\mathrm{s}})\), found from Eq.~\eqref{eq:pred_intensity} and then also normalised to \([0,1]\) by division by its maximum, the inner objective minimised by gradient based optimisation is
\begin{equation}
\mathcal{L}_{ij} =
\mathcal{L}_{\mathrm{H}}\left(I_{ij},\hat{I}_{ij};\delta_{ij}\right)
+ \lambda_{\mathrm{TV},ij}\Bigl(\mathrm{TV}(a_{ij})+\mathrm{TV}(\phi_{ij})\Bigr)
+ \lambda_{\mathrm{B},ij}\mathcal{L}_{\mathrm{B}}(a_{ij})
+ \lambda_{\mathrm{C},ij}\mathcal{L}_{\mathrm{C}}(a_{ij}),
\label{eq:inner_objective}
\end{equation}
where \(\boldsymbol{\theta}_{ij}\) (given by Eq.~\eqref{eq:policy_map}) specifies \(\delta_{ij}\), \(\lambda_{\mathrm{TV},ij}\), \(\lambda_{\mathrm{B},ij}\), \(\lambda_{\mathrm{C},ij}\), and the optimiser step sizes. Here \(\mathrm{TV}(\cdot)\) denotes the total variation functional, while \(\mathcal{L}_{\mathrm{B}}(\cdot)\) and \(\mathcal{L}_{\mathrm{C}}(\cdot)\) denote amplitude binarisation and contrast terms, respectively, as defined in the Methods. The data fidelity term, $\mathcal{L}_{\mathrm{H}}$,  is the Huber loss computed directly from the measured detector intensity and the predicted detector intensity.
\begin{equation}
\mathcal{L}_{\mathrm{H}}\left(I,\hat{I};\delta\right)
=
\frac{1}{|\Omega|}
\sum_{(x_{\mathrm{s}},y_{\mathrm{s}})\in\Omega}\!\!\!\!
\rho_{\delta}\left(I(x_{\mathrm{s}},y_{\mathrm{s}})-\hat{I}(x_{\mathrm{s}},y_{\mathrm{s}})\right),
\quad
\rho_{\delta}(r)=
\begin{cases}
\frac{1}{2}r^{2}, & |r|\le \delta, \\
\delta\bigl(|r|-\frac{1}{2}\delta\bigr), & |r|>\delta,
\end{cases}
\label{eq:huber}
\end{equation}
where \(\Omega\) denotes the set of detector pixels in the tile. Finally, the tile reconstructions are stitched to form full field amplitude and phase estimates, as described in the Methods Section.

A quantitative evaluation is performed to verify physics consistency at the detector plane. In addition to the Huber loss in Eq.~\eqref{eq:huber}, the mean squared error is computed as
\begin{equation}
\mathrm{MSE}\left(I,\hat{I}\right)
=
\frac{1}{|\Omega|}
\sum_{(x_{\mathrm{s}},y_{\mathrm{s}})\in\Omega}
\bigl(I(x_{\mathrm{s}},y_{\mathrm{s}})-\hat{I}(x_{\mathrm{s}},y_{\mathrm{s}})\bigr)^{2} \ge 0.
\label{eq:mse}
\end{equation}
Structural similarity is computed as \(\mathrm{SSIM}(I,\hat{I})\), with \(\mathrm{SSIM}\in[-1,1]\) and higher values indicating greater structural similarity. These metrics are computed per tile between the normalised measured intensity \(I_{ij}\) and the normalised predicted intensity \(\hat{I}_{ij}\) from Eq.~\eqref{eq:pred_intensity}, then averaged across all \(16\) tiles to report a mean and standard deviation.

\begin{figure}[t]
    \centering
    \includegraphics[width=\linewidth]{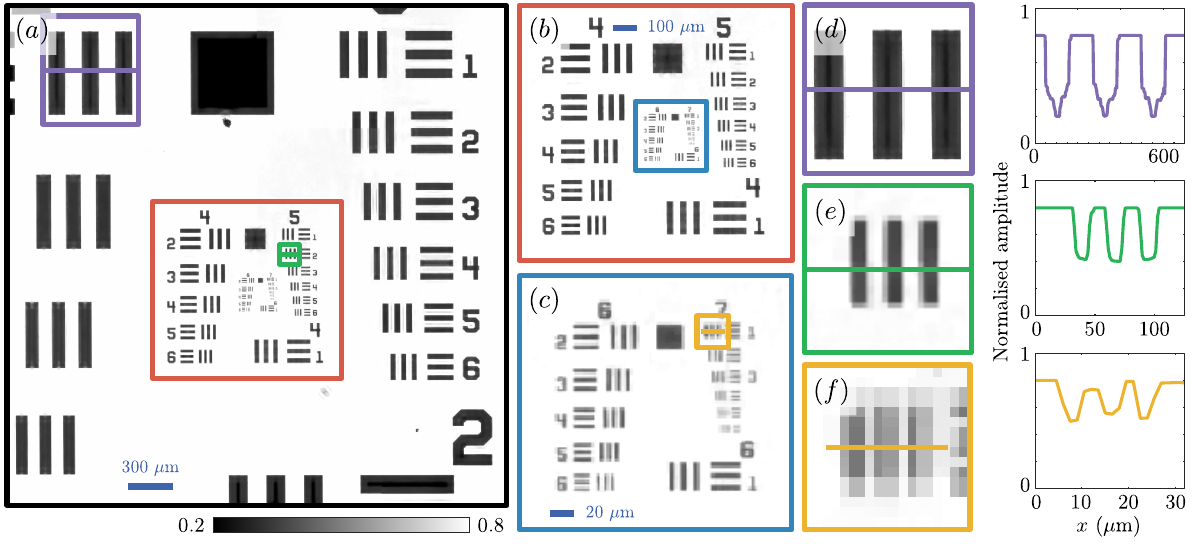}
    \caption{\textbf{Single-shot amplitude reconstruction of a resolution target.} (a) Reconstructed amplitude over the full field-of-view obtained from a single intensity measurement. (b) and (c) Coloured boxes indicate regions selected for magnified inspection. Enlarged views of the highlighted regions demonstrate recovery of fine spatial features across multiple resolution groups. (d)-(f) Line profiles extracted from the reconstructed amplitude along the indicated directions illustrate contrast preservation and spatial fidelity.}
    \label{fig:USAF}
\end{figure}

As an initial demonstration, the policy underlying the proposed reconstruction framework is evolved from a single recorded intensity measurement of a USAF resolution target and is first evaluated through reconstruction of that object. The reconstructed amplitude, along with representative line profiles, are shown in Fig.~\ref{fig:USAF}. Detector plane agreement is found to be consistently high across the full field-of-view, with mean Huber loss \((7.49 \pm 0.55)\times 10^{-5}\), mean squared error \((2.59 \pm 0.22)\times 10^{-4}\), and \(\mathrm{SSIM}=0.897 \pm 0.008\). These values indicate that the recovered complex object, when propagated through the forward model in Eq.~\eqref{eq:pred_intensity}, reproduces the recorded measurement with high fidelity over all tiles.

\begin{figure}[t]
    \centering
    \includegraphics[width=\linewidth]{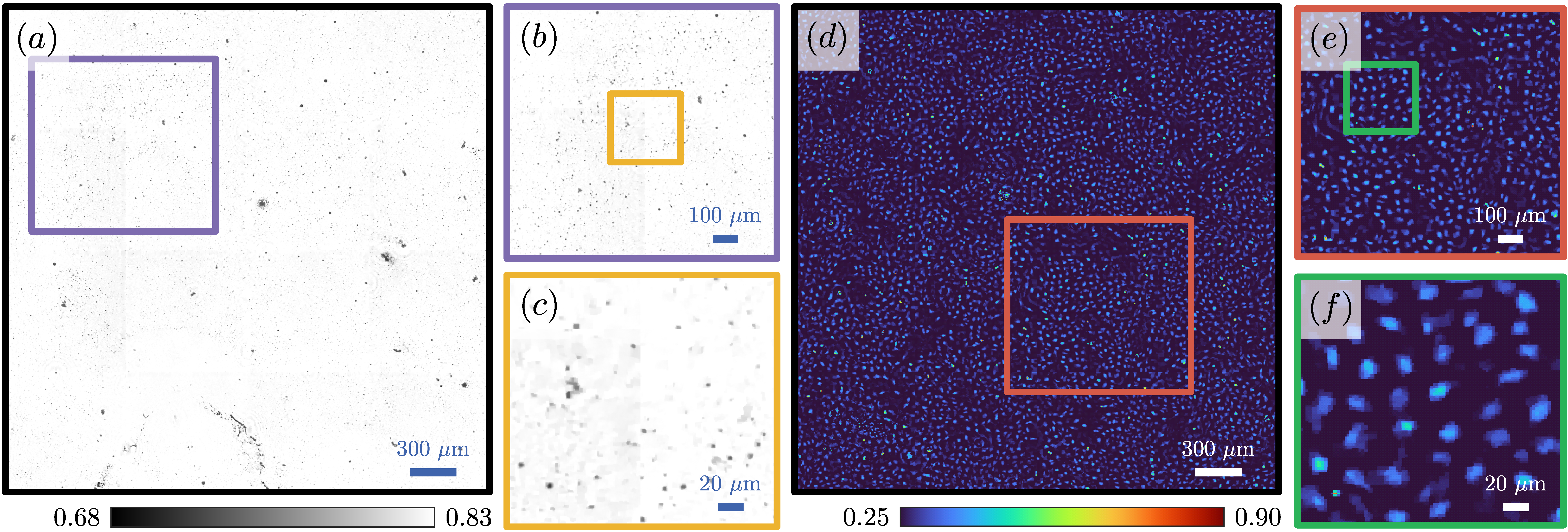}
    \caption{\textbf{Single-shot reconstructions of amplitude and phase dominant samples using evolved policy.} (a) Reconstructed amplitude of a microparticle sample at \(520~\mathrm{nm}\). (b) Inset from (a) shows representative regions with well-defined particle boundaries. (c) Inset from (b). (d) Reconstructed phase of a cellular sample at \(638~\mathrm{nm}\), shown after linear normalisation for visualisation. (e) Inset from (d) highlights cell boundaries and the separation of neighbouring cells. (f) Inset from (e).}
    \label{fig:micro_cells}
\end{figure}

The same evolved policy is then applied without modification to two distinct measurements that differ in structure and sample illumination wavelength. The measurement (at \(520~\mathrm{nm}\)) of randomly dispersed microparticles yields reconstructed amplitude maps with uniform contrast and well-defined particle boundaries, as shown in Fig.~\ref{fig:micro_cells}(a), Fig.~\ref{fig:micro_cells}(b), and Fig.~\ref{fig:micro_cells}(c). Detector plane consistency across all tiles remains high, with a mean Huber loss of \((6.85 \pm 1.54)\times 10^{-5}\), mean squared error of \((7.03 \pm 2.82)\times 10^{-4}\), and \(\mathrm{SSIM}=0.886 \pm 0.030\). For the phase-dominated cellular measurement at \(638~\mathrm{nm}\), the reconstructed phase map delineates cell boundaries and separates neighbouring cells, as shown in Fig.~\ref{fig:micro_cells}(d), Fig.~\ref{fig:micro_cells}(e), and Fig.~\ref{fig:micro_cells}(f). Fine intracellular structure is not resolved in this configuration because the system operates without magnification over a wide field-of-view and is therefore limited by sensor sampling and the effective numerical aperture set by the propagation geometry. With \(d_x = 1.4~\mu\mathrm{m}\) and a field-of-view of approximately \(3.36~\mathrm{mm}\), the combination of pixel-limited spatial frequency support, defocus sensitivity, and the phase-to-intensity ambiguity of inline measurements limits recovery of subcellular detail under a single-shot constraint. Despite this limitation, detector plane agreement remains strong, with mean Huber loss, mean square error and SSIM of \((1.20 \pm 0.11)\times 10^{-5}\),  \((6.25 \pm 1.10)\times 10^{-4}\) and \(0.867 \pm 0.017\) respectively.

Overall, a single evolved outer policy configures the inner solver to achieve stable convergence and high detector plane consistency across a resolution target, microparticles, and a phase-dominated biological specimen, without manual hyperparameter tuning, multiple measurements, or retraining. 

\subsection{Amyloid-Based Digital Bead Assay}
Robust single-shot lensless reconstruction is a practical necessity for digital assays, where the statistical confidence needed for diagnosis depends on consistent quantitative results over a large area and with tolerance to modest defocus and variability in sample positioning.
 To demonstrate this capability, the reconstruction framework developed above is integrated with an amyloid-based digital bead assay that tracks the loss of immobilised micro-beads as a proxy for amyloid dissociation kinetics \cite{RadhikaDigitalBeadEPPS}. In this assay, amyloid protofibrils are tethered onto latex beads which are subsequently immobilized on a glass substrate (see Methods for details). Upon introduction of a known disaggregation agent, specifically EPPS (4-(2-hydroxyethyl)-1-piperazinepropanesulfonic acid), subsequent destabilisation of the amyloid structures leads to bead release from the substrate, enabling time-resolved quantification of amyloid dissociation kinetics \cite{Kim2015EPPS,Nair2026.02.22.706957} by counting the remaining bound beads. The digital profiling protocol follows the workflow reported in \cite{RadhikaDigitalBeadEPPS}. 

\begin{figure*}[t]
    \centering
    \includegraphics[width=\linewidth]{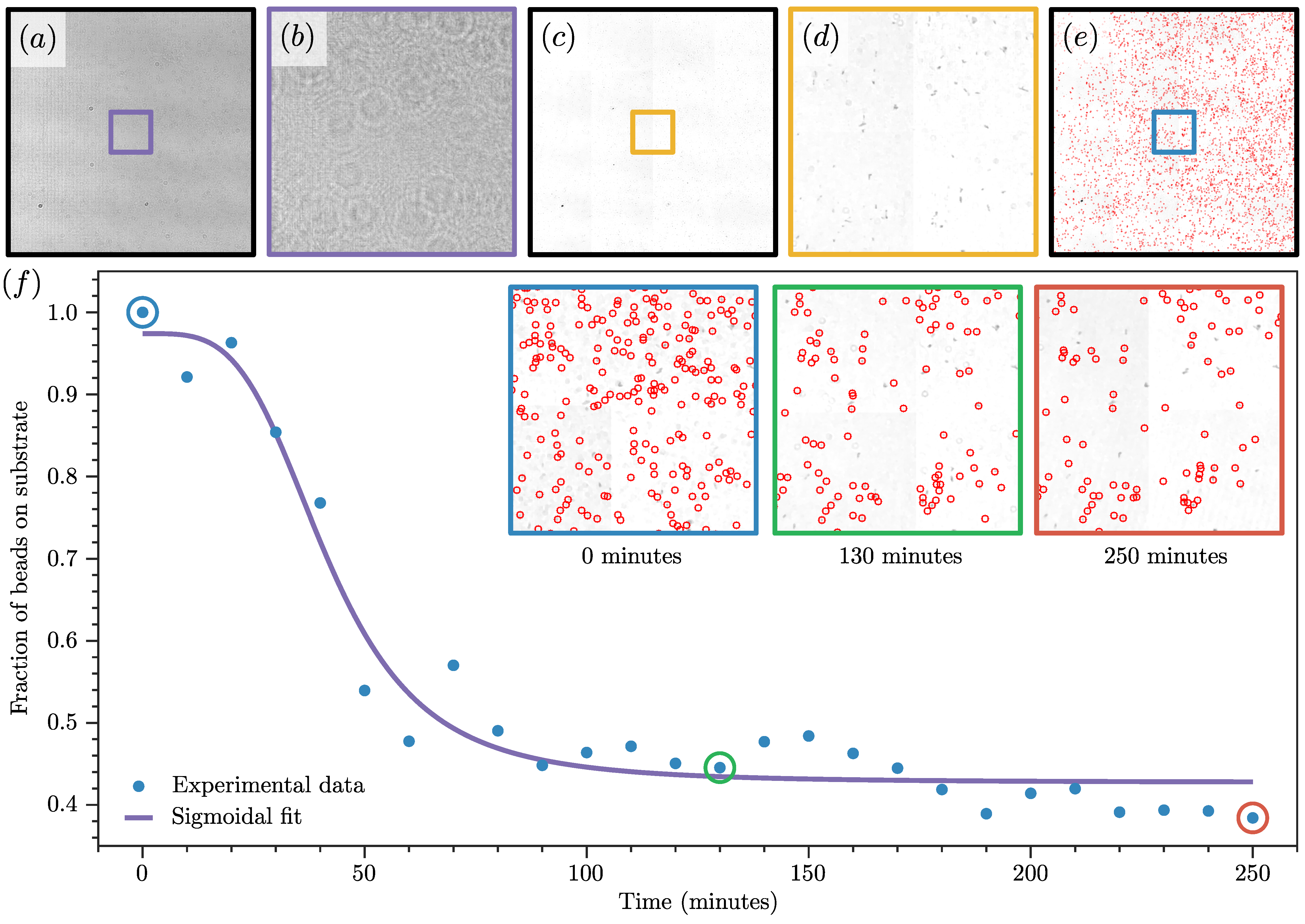}
    \caption{\textbf{Digital bead profiling using single-shot lensless reconstruction.} (a) Full field lensless measurement of the amyloid-based bead assay at \(T = 60\) minutes. (b) Magnified inset of the region highlighted in (a), showing the raw lensless measurement before reconstruction. (c) Reconstructed full-field object-plane amplitude image corresponding to the measurement in (a). (d) Magnified inset of the region highlighted in (c). (e) Accepted bead detections obtained from the reconstructed image in (c). Candidate bead centres are first identified from the reconstructed amplitude image and are then filtered using the analysis mask described in the Methods Section to reject image artefacts and non-bead structures. (f) Dissociation kinetics of the assay, shown as the normalised fraction of beads remaining on the substrate as a function of time, together with an inverse sigmoid fit. The inset images in (f) show representative reconstructed regions and their corresponding accepted bead detections at selected time points, taken from the region corresponding to the inset views in ((e).}
\label{fig:AmyloidAssay}
\end{figure*}

A time-ordered sequence of assay frames is acquired under the same inline lensless configuration used for amplitude-dominant samples. Each frame is reconstructed using the same evolved policy and inner solver described above and in the Methods Section, yielding a sequence of object-plane amplitude images. Bead counting and spatial profiling are then performed on the reconstructed amplitude sequence using the analysis pipeline described in the Methods Section. Figure~\ref{fig:AmyloidAssay} summarises the full-field reconstruction and the downstream digital readout. The full field lensless measurement of the assay is shown in Fig.~\ref{fig:AmyloidAssay}(a), with a representative inset region highlighted for closer inspection. The corresponding inset from the raw lensless measurement is shown in Fig.~\ref{fig:AmyloidAssay}(b). The reconstructed full-field amplitude image is shown in Fig.~\ref{fig:AmyloidAssay}(c), and the corresponding reconstructed inset is shown in Fig.~\ref{fig:AmyloidAssay}(d). Candidate bead centres are then detected from the reconstructed image and subjected to the filtering procedure described in the Methods Section to reject artefacts and non-bead structures. The resulting accepted bead detections are shown in Fig.~\ref{fig:AmyloidAssay}(e). Figure~\ref{fig:AmyloidAssay}(f) shows the dissociation kinetics, quantified as the fraction of beads remaining on the substrate as a function of time, together with representative reconstructed inset regions and their corresponding accepted bead detections at selected time points.

Temporal kinetics are quantified using the frame-level bead count \(N_k\) and the normalised fraction of beads remaining on the substrate
\begin{equation}
f_k=\frac{N_k}{N_0},
\label{eqBeadFraction}
\end{equation}
where \(k\) indexes acquisition time. In this experiment, \(N_0=9453
\) beads are detected at the initial time point and \(N_{25}=3631
\) beads are detected at the final time point at \(250\) minutes, corresponding to \(f_{25}=0.3841\). To quantitatively characterise the dissociation dynamics, the temporal evolution of $f_k$ is fitted using a sigmoidal Hill-type decay function:
\begin{equation}
f_k = f_{\mathrm{final}} + \frac{f_0 - f_{\mathrm{final}}}{1 + \left(\frac{t}{t_{1/2}}\right)^n},
\end{equation}
where $f_0$ and $f_{\mathrm{final}}$ represent the initial and final fraction of substrate-bound beads, respectively, $t_{1/2}$ denotes the characteristic half-time corresponding to 50\% bead loss, and $n$ is the Hill coefficient describing the steepness of the transition. This formulation captures the non-linear dissociation behaviour associated with amyloid disaggregation and enables extraction of key kinetic parameters, including the dissociation half-time, the effective rate of bead loss (from the slope at $t_{1/2}$), the degree of cooperativity ($n$), and the residual bound fraction ($f_{\mathrm{final}}/f_0$). These parameters provide a compact quantitative description of the disaggregation process and facilitate comparison across experimental conditions. Kinetic analysis based on the Hill model yielded a $k_D$ value of $40.16 \pm 0.00416~\mathrm{min}$, corresponding to the half-time ($t_{1/2}$), and a Hill coefficient ($n$) of $3.816$, characterising the amyloid disaggregation process. These results match the kinetic trends reported in \cite{RadhikaDigitalBeadEPPS} while leveraging a millimetre scale field-of-view, enabling digital profiling over a substantially larger assay area without sacrificing single-shot acquisition.

\section{Discussion}
\label{secDiscussion}

Single-shot lensless imaging is fundamentally constrained by limited measurement diversity, which makes reconstruction highly sensitive to defocus, interference artefacts, and optimisation choices. The central aim of this work is therefore not only to enforce a wave-optics forward model, but also to establish a reconstruction procedure that remains stable when the sample class and acquisition conditions change. The results indicate that automatic configuration of an iterative solver can convert a fragile, hand-tuned pipeline into a repeatable procedure that operates directly on the recorded intensity and jointly estimates amplitude, phase, and propagation distance.

A key feature of the proposed framework is the separation of policy evolution from per-measurement inference. Genetic programming explores a space of bounded hyperparameter mappings, allowing the solver schedule to adapt to measurable properties of the recorded diffraction pattern rather than relying on a fixed manually selected schedule. This design directly targets a major practical source of failure in single-shot lensless image reconstruction, namely unstable iterative updates and an incorrect balance of regularisation under varying contrast and defocus. Hard bounding of the propagation distance during optimisation further suppresses physically implausible solutions and reduces sensitivity to initialisation. Tiling complements this strategy by localising the optimisation problem, which is appropriate when fringe density and effective defocus vary across a wide field of view.

\begin{figure*}[t]
    \centering
    \includegraphics[width=\linewidth]{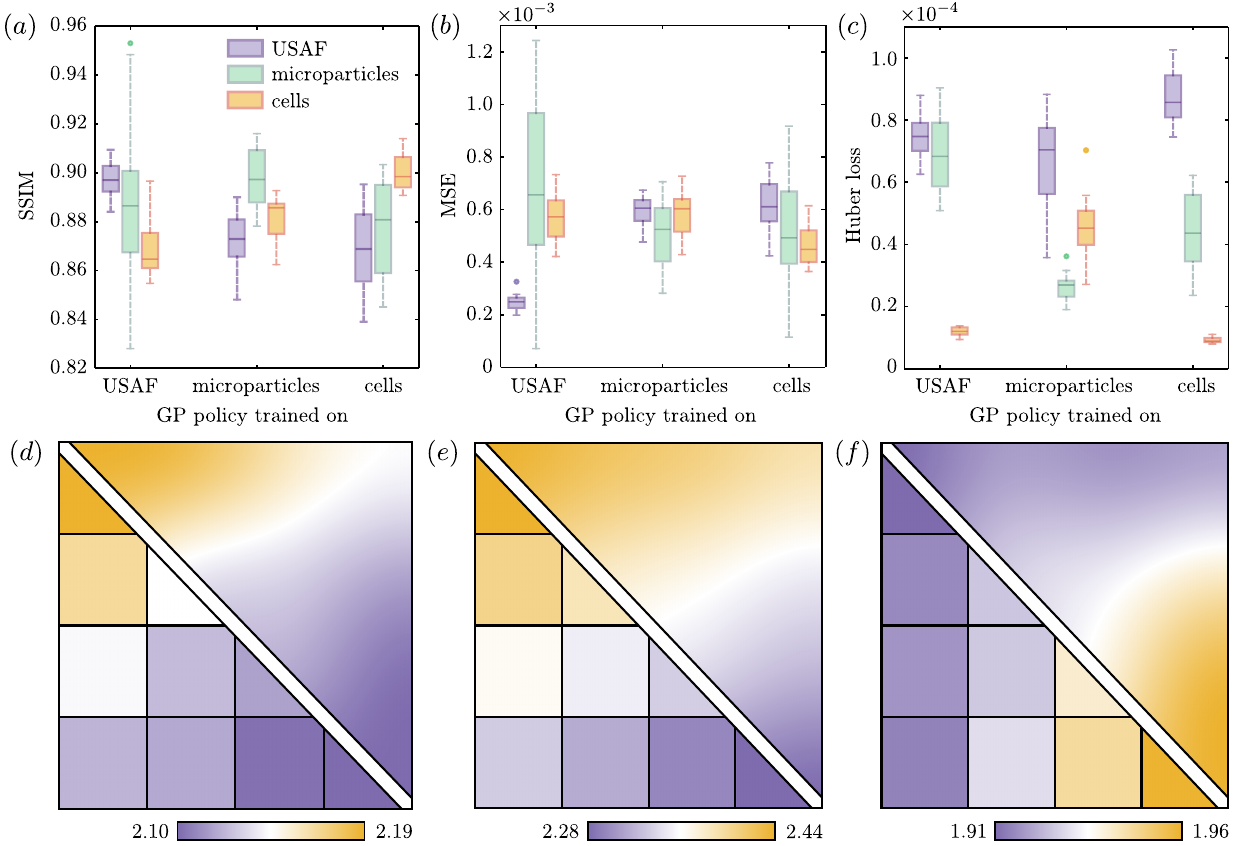}
    \caption{
\textbf{Performance generalisation and tile-wise propagation distance estimation for genetic programming-based policies.}
(a) Structural similarity index (b) mean square error and (c) Huber loss between the predicted detector plane intensity and the measured intensity when the policy is evolved using a USAF resolution target, a microparticle sample, or a cellular sample and then tested across all object types.
(d) Final tile-wise propagation-distance estimates for the USAF target, (e) dispersed microparticles and (f) U2OS cells shown using a diagonal split representation. The upper-right triangle displays the spatially interpolated distance map on the full sensor grid, whereas the lower-left triangle displays the original 4×4 tile-wise estimates.}
    \label{GeneralizationStats}
\end{figure*}

The observed generalisation across object classes is best interpreted in terms of measurement statistics. The evolved mapping uses low-order descriptors and tile location to set step sizes, regularisation weights, and stopping behaviour. These descriptors capture features that influence the optimisation landscape, including local contrast and spatial non-uniformity, while remaining agnostic to specific object content. As a result, a policy evolved on one sample can remain effective on previously unseen object classes because the controlled variables are coupled to the physics of intensity formation and to solver stability, rather than to class-specific appearance. This behaviour is consistent with Fig.~\ref{GeneralizationStats}(a)--(c), where policies evolved on the USAF target, microparticle sample, or cellular sample continue to produce strong detector-plane agreement when transferred across the other object classes. The variation across test cases indicates some dependence on the measurement statistics of the evolution sample, but the absence of severe performance collapse under cross-sample transfer supports the claim that the learned policy is not narrowly tied to a single training object.

The tile-wise propagation-distance estimates provide an additional source of robustness that is particularly relevant for field-deployable operation. Residual axial offsets can arise from mild sample tilt, imperfect placement, or spatial non-uniformity across the field, and these mismatches are among the fastest routes to degradation in inline diffraction measurements. Estimating distance on a tile-wise basis therefore acts as a computational correction for these effects while relaxing mechanical tolerances in compact hardware. This behaviour is visualised in Fig.~\ref{GeneralizationStats}(d)--(f), where the final per-tile distance estimates are spatially interpolated onto the full sensor grid. The resulting maps provide a compact summary of axial variation across the field of view and help explain why stable reconstruction can be maintained even when a single global distance would be insufficient.

These properties are especially important in the amyloid-based digital bead assay, where the readout is defined by the number of beads remaining on the substrate over time. In conventional dark-field or bright-field implementations, the restricted field of view typically necessitates acquisition of multiple overlapping regions followed by image stitching, which increases experimental complexity, extends processing time, and can introduce variability across acquisitions. In the present framework, a single lensless measurement covers the entire assay area, and reconstruction restores the spatial information needed for bead detection and counting throughout the field. This enables direct wide-field quantification without sequential scanning or stitching. At the same time, the reconstruction stage is essential, because raw lensless intensity patterns are shaped by diffraction, defocus, and overlap between neighbouring bead signatures, which limits reliable direct quantification from the sensor measurement alone. By recovering object-plane information and estimating the effective propagation distance, the proposed approach enables consistent bead detection across varying experimental conditions, thereby supporting scalable quantitative analysis of amyloid disaggregation kinetics.

The current findings, however, also point to specific paths for future advancement. Due to pixel-limited sampling and the lack of additional measurement diversity, fine intracellular phase features are not fully recovered under the current field-of-view conditions without magnification. Similarly, repeated tile-wise optimization continues to dominate the computational cost. Nonetheless, by extending the forward model and the policy inputs, more measurement diversity could be introduced. Warm starts, parallel execution across tiles, mixed-precision arithmetic, and more aggressive early-stopping criteria that remain aligned with detector-plane consistency could all shorten reconstruction times without changing the underlying physics. Together, these observations indicate that the existing framework is already effective for large-field-of-view quantitative lensless readout and provides a clear and promising path toward richer phase recovery, faster operation, and wider deployment across practical assay and imaging settings.

\section{Methods}
\label{secMethods}

\subsection{Preparation of the Amyloid-Based Digital Bead Assay}
\label{AmyloidAssay_rad}
Carboxylated latex beads (average diameter 2~$\mu$m, 2.5~wt\%) were obtained from Sigma-Aldrich (904465-2G, Sigma-Aldrich). Human $\beta$-amyloid peptide A$\beta$(1--42) was purchased from Abcam. Surface functionalisation of latex beads with A$\beta$ was carried out using 1-Ethyl-3-(3-dimethylaminopropyl)carbodiimide to N-hydroxysuccinimide (EDC-NHS) carbodiimide coupling chemistry.

Initially, carboxylated latex beads were diluted in phosphate-buffered saline (PBS), thoroughly mixed, and centrifuged to remove any impurities. Following removal of the supernatant, the beads were resuspended in freshly prepared carboxyl activation solution containing EDC and NHS mixed in a 2:1 mass ratio and dissolved in 0.1~M MES buffer prepared with reverse osmosis water. The suspension was gently mixed to activate surface carboxyl groups.

Subsequently, one-hour pre-incubated A$\beta$(1--42) solution ($\sim 1~\mu$M) was introduced into the activated bead mixture, followed by addition of 50~$\mu$l Trisma-hydrogen chloride buffer to facilitate coupling. The reaction mixture was then centrifuged again at 16000 rpm and the supernatant was discarded to remove unbound components.

To block remaining reactive sites and stabilise the suspension, 2 ml of 0.1\% (w/w) polyvinylpyrrolidone (PVP) solution was added. Finally, the beads were washed and resuspended in PBS, and the resulting A$\beta$-coated latex beads were stored at 4$^\circ$C until further use.

To prepare the digital counting assay for quantifying A$\beta$ dissociation kinetics, glass substrates were functionalised to support stable immobilisation of amyloid-coated latex beads. The substrates were first thoroughly cleaned with an organic solvent, rinsed with deionised water, and dried. A thin layer of poly-L-lysine was applied to introduce primary amines, creating an amine-rich surface suitable for covalent coupling.

A$\beta$-coated latex beads were then immobilised on the poly-L-lysine--treated surface using EDC-NHS carbodiimide chemistry, ensuring stable covalent attachment. Following immobilisation, excess beads were removed through gentle washing. To verify attachment stability, substrates underwent repeated washing cycles, and microscopic imaging confirmed that the majority of beads remained firmly adhered with minimal clustering. The resulting surfaces exhibited a uniform distribution of individual A$\beta$-coated beads and were immediately used for dissociation experiments.

\subsection{Experimental Details of the Lensless Imaging Setup}
\label{ExpDet}
The lensless imaging system uses a laser as an illumination source (Cobolt 06-MLD 520 nm and 638 nm) and a camera (Raspberry pi camera module 3, Sony  IMX708, resolution: 11.9 megapixels, sensor size: 7.4 mm diagonal, pixel size: $1.4~\mu\mbox{m} \times 1.4~\mu$m) as detector, with a total separation of approximately 0.5 m. The illumination is not collimated and the beam expands significantly before reaching the sample, ensuring a uniform field of illumination. A 3D printed sample holder allows the coverslip to be mounted to a distance less than 5 mm of the camera sensor.

We mixed 10 $\mu$l of the 2 $\mu$m silica beads (904465-2G, Sigma-Aldrich) with 20 $\mu$l water (MilliQ). The solution was spin-coated (30 $\mu$l, 1000 rpm, 500 rpm/s, 2 min) on to a coverslip to create individually separated particles dispersed on the surface. For the human osteosarcoma cell sample, U2OS cells were grown in Fibronectin coated surfaces. The cells were seeded for 12-16 hours, after which they were harvested and deposited directly between two coverslips. Paraformaldehyde was used to fix the cells for imaging.

\subsection{Image Reconstruction Pipeline}
\label{subsecImageReconstructionPipeline}

A single lensless intensity measurement is recorded at the sensor \cite{Ozcan2016Lensless,Boominathan2022OpticaReview,rosen2024roadmap}. The reconstruction task is formulated as the estimation of a complex transmission coefficient  in the object plane (assuming uniform illumination) from a single measured intensity image under a scalar diffraction model \cite{Shechtman2015PR,Fienup1982PhaseRetrieval,Gerchberg1972GS}. The pipeline has two coupled components. An inner physics-constrained optimisation estimates tile-wise amplitude, tile-wise phase, and tile-wise object-sensor propagation distance. An outer genetic programming stage evolves a single global rule that converts simple measurement descriptors into the hyperparameters required by the inner optimiser. The evolved rule is learned once from a single training measurement and is then reused for subsequent measurements acquired under the same optical configuration, with external acquisition parameters such as wavelength updated as required.

Let $\mathbf{r}_{\mathrm{o}}=(x_{\mathrm{o}},y_{\mathrm{o}})$ denote transverse coordinates in the object plane and let $\mathbf{r}_{\mathrm{s}}=(x_{\mathrm{s}},y_{\mathrm{s}})$ denote transverse coordinates in the sensor plane. The object is modelled as a complex transmittance
\begin{equation}
\tau(\mathbf{r}_{\mathrm{o}}) = a(\mathbf{r}_{\mathrm{o}})\,\exp\bigl(\mathrm{i}\,\phi(\mathbf{r}_{\mathrm{o}})\bigr),
\label{eqObjectModelPipe}
\end{equation}
where $a(\mathbf{r}_{\mathrm{o}})\in[0,1]$ is the amplitude transmission coefficient and $\phi(\mathbf{r}_{\mathrm{o}})\in\mathbb{R}$ is object-induced phase delay. The propagated field at the sensor plane is
\begin{equation}
v(\mathbf{r}_{\mathrm{s}};z) = \mathcal{P}_{z}\left[\tau\right](\mathbf{r}_{\mathrm{s}}),
\label{eqForwardFieldPipe}
\end{equation}
and the predicted sensor intensity is
\begin{equation}
I^{\mathrm{pred}}(\mathbf{r}_{\mathrm{s}};z) = \left|v(\mathbf{r}_{\mathrm{s}};z)\right|^{2}.
\label{eqForwardIntensityPipe}
\end{equation}
The operator $\mathcal{P}_{z}$ is implemented using a Fourier domain Fresnel propagator \cite{Goodman2005Fourier,Voelz2011ComputationalFourierOptics}
\begin{equation}
\mathcal{P}_{z}[\tau] =
\mathcal{F}^{-1}\left\{
\mathcal{F}\{\tau\}\,H(f_x,f_y;z)
\right\},
\label{eqFresnelPropPipe}
\end{equation}
with transfer function
\begin{equation}
H(f_x,f_y;z) =
\exp\Bigl(
-\mathrm{i}\,\pi \lambda z \bigl(f_x^{2}+f_y^{2}\bigr)
\Bigr),
\label{eqFresnelTFPipe}
\end{equation}
where $\mathcal{F}$ denotes the two-dimensional Fourier transform, $(f_x,f_y)$ are spatial frequency coordinates determined by the sampling pitch, and $\lambda$ is the illumination wavelength. Measured and predicted intensities are normalised to \([0,1]\) before computing the objective terms used for optimisation and evaluation.

The full sensor image is partitioned into a fixed grid of tiles. The inner solver reconstructs $(a,\phi,z)$ independently for each tile. The reconstructed tile cores are then stitched into a full-field mosaic by inserting each tile core at its predefined spatial location.

\subsubsection{Inner Tile Reconstruction by Gradient-Based Optimisation}
\label{subsubsecInnerTileGD}

For a tile indexed by \((i,j)\) with detector-pixel set \(\Omega_{ij}\), the unknowns are the tile amplitude \(a_{ij}\), tile phase \(\phi_{ij}\), and tile propagation distance \(z_{ij}\) between the object and the sensor. The distance \(z_{ij}\) is optimised using a bounded reparameterisation in terms of an unconstrained variable \(\zeta_{ij}\in\mathbb{R}\),
\begin{equation}
z_{ij} =
z_{\min} + (z_{\max}-z_{\min})\, \sigma(\zeta_{ij}),
\label{eqZReparamPipe}
\end{equation}
where \(\sigma(\cdot)\) is the logistic function. This construction guarantees \(z_{ij}\in[z_{\min},z_{\max}]\) throughout optimisation and prevents physically implausible axial updates. Amplitude feasibility is enforced by projection onto \([0,1]\) after each update. Phase is treated as an unbounded real-valued variable during optimisation. For visualisation, the reconstructed phase can be wrapped to a principal interval and linearly scaled to \([0,1]\) for display, while the optimisation itself is performed on the underlying unbounded parameter.

Let \(I_{ij}(x_{\mathrm{s}},y_{\mathrm{s}})\) denote the measured tile intensity on the sensor plane after normalisation, and let \(\hat{I}_{ij}(x_{\mathrm{s}},y_{\mathrm{s}})\) denote the predicted intensity computed from Eqs.~\eqref{eqForwardFieldPipe} and \eqref{eqForwardIntensityPipe} using the current tile estimates of amplitude, phase, and distance. The tile objective is
\begin{equation}
\mathcal{L}_{ij}
=
\mathcal{L}_{\mathrm{H}}\left(I_{ij},\hat{I}_{ij};\delta_{ij}\right)
+
\lambda_{\mathrm{TV},ij}\mathcal{L}_{\mathrm{TV},ij}
+
\lambda_{\mathrm{B},ij}\mathcal{L}_{\mathrm{B},ij}
+
\lambda_{\mathrm{C},ij}\mathcal{L}_{\mathrm{C},ij},
\label{eqTileObjectivePipe}
\end{equation}
where \(\delta_{ij}\), \(\lambda_{\mathrm{TV},ij}\), \(\lambda_{\mathrm{B},ij}\), and \(\lambda_{\mathrm{C},ij}\) are supplied by the outer policy for each tile. The data fidelity term uses a Huber penalty \cite{Huber1964Robust}
\begin{equation}
\mathcal{L}_{\mathrm{H}}\left(I,\hat{I};\delta\right)
=
\frac{1}{|\Omega_{ij}|}
\sum_{(x_{\mathrm{s}},y_{\mathrm{s}})\in\Omega_{ij}}
\rho_{\delta}\left(I(x_{\mathrm{s}},y_{\mathrm{s}})-\hat{I}(x_{\mathrm{s}},y_{\mathrm{s}})\right),
\label{eqHuberDataPipe}
\end{equation}
with
\begin{equation}
\rho_{\delta}(r)=
\begin{cases}
\frac{1}{2}r^{2}, & |r|\le \delta,\\
\delta\Bigl(|r|-\frac{1}{2}\delta\Bigr), & |r|>\delta.
\end{cases}
\label{eqHuberPenaltyPipe}
\end{equation}
The threshold \(\delta\) controls the transition between a quadratic penalty for small residuals and a linear penalty for large residuals. In this pipeline, \(\delta_{ij}\) is not manually chosen, but is selected automatically by the outer policy from a predefined bounded interval. Larger values behave more like least squares, whereas smaller values reduce the influence of outliers and decrease sensitivity to localised mismatch.

Total variation regularisation promotes piecewise smooth amplitude and phase and suppresses high-spatial-frequency artefacts \cite{Rudin1992ROF}. Using anisotropic forward differences for a sampled function \(f\), \(\Delta_x f(m,n)=f(m,n{+}1)-f(m,n)\) and \(\Delta_y f(m,n)=f(m{+}1,n)-f(m,n)\), the tile total variation is
\begin{equation}
\mathrm{TV}(f) =
\frac{1}{|\Omega_{ij}|}
\sum_{(m,n)\in\Omega_{ij}}
\Bigl(
|\Delta_x f(m,n)| + |\Delta_y f(m,n)|
\Bigr),
\label{eqTVDefPipe}
\end{equation}
and the corresponding smoothness term is
\begin{equation}
\mathcal{L}_{\mathrm{TV},ij} =
\mathrm{TV}(a_{ij}) + \mathrm{TV}(\phi_{ij}).
\label{eqTVRegPipe}
\end{equation}

A binarisation term is included for amplitude objects because two of the experimental object classes are well approximated by sparse opaque structures on a transmissive background, which is consistent with amplitude values concentrating near \(0\) and \(1\). The penalty is
\begin{equation}
\mathcal{L}_{\mathrm{B},ij} =
\frac{1}{|\Omega_{ij}|}
\sum_{(x_{\mathrm{o}},y_{\mathrm{o}})\in\Omega_{ij}}
4\,a_{ij}(x_{\mathrm{o}},y_{\mathrm{o}})
\Bigl(1-a_{ij}(x_{\mathrm{o}},y_{\mathrm{o}})\Bigr),
\label{eqBinaryRegPipe}
\end{equation}
where the weight \(\lambda_{\mathrm{B},ij}\) is selected by the outer policy from a narrow interval that includes values close to zero, allowing the contribution to become negligible for phase-dominant measurements where an explicit binarisation bias is less appropriate.

A contrast term based on the tile amplitude standard deviation is
\begin{equation}
\mathcal{L}_{\mathrm{C},ij} = \mathrm{Std}\bigl(a_{ij}\bigr),
\label{eqContrastRegPipe}
\end{equation}
where \(\mathrm{Std}(a_{ij})\) is computed over all pixels in the tile. The sign and magnitude of \(\lambda_{\mathrm{C},ij}\) determine whether amplitude contrast is encouraged or discouraged within the objective.

Optimisation of Eq.~\eqref{eqTileObjectivePipe} is performed by automatic differentiation through the forward model in Eqs.~\eqref{eqForwardFieldPipe} and \eqref{eqFresnelPropPipe}. An adaptive first-order optimiser is used with separate learning rates for the amplitude and phase block and for the distance variable \cite{Kingma2015Adam}.

\subsubsection{Outer Policy Evolution by Genetic Programming}
\label{subsubsecOuterPolicyGP}

The outer stage searches for a single global rule that sets tile-specific hyperparameters for the inner optimiser. Genetic programming is used to represent this rule as a set of symbolic expression trees that transform a small number of descriptors of the measured intensity into hyperparameters \cite{Koza1992GP,Poli2008FieldGuideGP,EibenSmith2015EC}. In this context, each tree is interpreted as a single closed-form function that outputs one hyperparameter. The full policy is therefore a bundle of \(K\) trees, with one tree assigned to each hyperparameter required by the inner optimisation.

For tile \((i,j)\), a feature vector \(\mathbf{f}_{ij}\in\mathbb{R}^{7}\) is computed from the measured intensity values
\begin{equation}
\mathbf{f}_{ij} =
\bigl[
X_{ij},\;
Y_{ij},\;
m_{ij},\;
s_{ij},\;
g_M,\;
g_S,\;
C
\bigr]^{\mathsf{T}}.
\label{eqTileFeaturesPipe}
\end{equation}
Here, \((X_{ij}, Y_{ij})\) denotes the normalised tile-centre coordinates within the full sensor image. The scalars \(m_{ij}\) and \(s_{ij}\) are the mean and standard deviation of the measured tile intensity values after normalisation, whereas \(g_M\) and \(g_S\) are the corresponding global statistics computed over the full measured intensity image. The term \(C\) is a fixed bias input, common to all tiles, that allows constant offsets in the evolved mappings. Including \((X_{ij}, Y_{ij})\) allows the policy to vary hyperparameters across the field of view, which is useful when residual tilt or spatially varying defocus produces location-dependent measurement structure. The local and global statistics provide compact descriptors of contrast and noise level that help the policy adjust step sizes and regularisation strengths without introducing additional latent variables.

Each policy contains \(K=8\) trees \(\{g_k\}_{k=1}^{K}\). For a given tile, the trees produce raw scalars \(r_{k,ij}=g_k(\mathbf{f}_{ij})\). Each raw output is squashed by a logistic function, \(s_{k,ij}=1/(1+\exp(-r_{k,ij}))\), and mapped to a bounded interval through \(\theta_{k,ij}=\ell_k+(u_k-\ell_k)s_{k,ij}\), where \((\ell_k,u_k)\) are predefined bounds. This bounded mapping guarantees that every candidate policy produces numerically admissible hyperparameters for every tile, even when the symbolic expressions are modified by recombination and random mutation during search.

The resulting hyperparameter vector for tile \((i,j)\) is
\begin{equation}
\boldsymbol{\theta}_{ij} =
\bigl[
n_{\mathrm{iter},ij},\;
\eta_{a\phi,ij},\;
\eta_{z,ij},\;
\lambda_{\mathrm{TV},ij},\;
\lambda_{\mathrm{B},ij},\;
\lambda_{\mathrm{C},ij},\;
\delta_{ij},\;
z_{0,ij}
\bigr]^{\mathsf{T}},
\label{eqHyperparameterVectorPipe}
\end{equation}
where \(n_{\mathrm{iter},ij}\) is the number of inner iterations, \(\eta_{a\phi,ij}\) and \(\eta_{z,ij}\) are the learning rates for the amplitude-phase block and the distance variable, \(\lambda_{\mathrm{TV},ij}\) is the total variation weight, \(\lambda_{\mathrm{B},ij}\) is the binarisation weight, \(\lambda_{\mathrm{C},ij}\) is the contrast weight, \(\delta_{ij}\) is the Huber threshold in Eqs.~\eqref{eqHuberDataPipe} and \eqref{eqHuberPenaltyPipe}, and \(z_{0,ij}\) initialises the distance variable that is later refined through Eq.~\eqref{eqZReparamPipe}. The hyperparameters are tile-specific, but the policy itself is global. It is evolved using the full measurement by evaluating every candidate policy across all tiles and aggregating performance into a single fitness score.

Evolution proceeds by maintaining a population of candidate policies. New candidates are produced by recombining subexpressions between two policies and by randomly altering a subexpression within one policy. Multiobjective selection is performed using nondominated sorting and crowding distance to balance reconstruction quality, computational cost, and symbolic complexity \cite{Deb2001MOEA,EibenSmith2015EC}. The fitness vector is
\begin{equation}
\mathbf{F} =
\bigl[
\overline{\mathrm{SSIM}},\;
T,\;
S
\bigr]^{\mathsf{T}},
\label{eqFitnessVectorPipe}
\end{equation}
where \(\overline{\mathrm{SSIM}}\) is the mean structural similarity index aggregated across tiles, \(T\) is evaluation time, and \(S\) is the total node count across the \(K\) trees. For each tile, the structural similarity index is computed as \(\mathrm{SSIM}(I_{ij},\hat{I}_{ij})\), where \(\hat{I}_{ij}\) is the predicted tile intensity obtained by propagating the reconstructed object field through Eqs.~\eqref{eqForwardFieldPipe} and \eqref{eqForwardIntensityPipe}. The mean score \(\overline{\mathrm{SSIM}}\) is then computed by averaging over all tiles. In this way, a per-tile physics-consistency measure is aggregated into a global objective that is used to evolve a single policy for later reuse across measurements.

\subsection{Data Analysis for Amyloid Based Digital Bead Assay}
\label{AnalData}

The digital bead assay readout is obtained from a time-ordered sequence of reconstructed object-plane amplitude images \(a_k(\mathbf{r}_{\mathrm{o}})\), where \(k\in\{1,\ldots,K\}\) indexes the acquisition frame and \(\mathbf{r}_{\mathrm{o}}=(x_{\mathrm{o}},y_{\mathrm{o}})\) denotes coordinates on the reconstructed grid. Each frame is normalised to \([0,1]\) using minimum and maximum scaling
\begin{equation}
a_k^{\mathrm{n}}(\mathbf{r}_{\mathrm{o}})=
\frac{a_k(\mathbf{r}_{\mathrm{o}})-\min_{\mathbf{r}_{\mathrm{o}}}a_k(\mathbf{r}_{\mathrm{o}})}
{\max_{\mathbf{r}_{\mathrm{o}}}a_k(\mathbf{r}_{\mathrm{o}})-\min_{\mathbf{r}_{\mathrm{o}}}a_k(\mathbf{r}_{\mathrm{o}})+\epsilon},
\label{eqAssayNorm}
\end{equation}
where \(\epsilon>0\) prevents division by zero. Since beads appear as dark structures in amplitude, detection is performed on the inverted image
\begin{equation}
I_k(\mathbf{r}_{\mathrm{o}})=1-a_k^{\mathrm{n}}(\mathbf{r}_{\mathrm{o}}).
\label{eqAssayInvert}
\end{equation}

If \(K>1\), translation-only registration is applied sequentially so that each frame is aligned to the preceding registered frame. The first frame defines the initial reference, so \(I^{\mathrm{reg}}_{1}(\mathbf{r}_{\mathrm{o}})=I_{1}(\mathbf{r}_{\mathrm{o}})\). For \(k\ge 2\), the registered frame is
\begin{equation}
I_k^{\mathrm{reg}}(\mathbf{r}_{\mathrm{o}})=I_k\bigl(\mathbf{r}_{\mathrm{o}}+\boldsymbol{\Delta}_k\bigr),
\label{eqAssayReg}
\end{equation}
where \(\boldsymbol{\Delta}_k=(\Delta x_k,\Delta y_k)\) is an integer translation estimated by normalised cross-correlation over several stable anchor regions, with the final frame translation taken as the componentwise median of the anchor-specific translations. After registration, the analysis domain is restricted to the maximal common overlap shared by all registered frames.

Each registered frame is denoised using a median filter. A smooth background is then estimated by Gaussian filtering with standard deviation \(\sigma_{\mathrm{BG}}\),
\begin{equation}
B_k(\mathbf{r}_{\mathrm{o}})=\bigl(G_{\sigma_{\mathrm{BG}}}*I_k^{\mathrm{reg}}\bigr)(\mathbf{r}_{\mathrm{o}}),
\label{eqAssayBG}
\end{equation}
and background-flattened contrast is obtained by subtraction with non-negativity enforcement
\begin{equation}
F_k(\mathbf{r}_{\mathrm{o}})=\max\bigl(0,\,I_k^{\mathrm{reg}}(\mathbf{r}_{\mathrm{o}})-B_k(\mathbf{r}_{\mathrm{o}})\bigr),
\label{eqAssayFlat}
\end{equation}
where \(G_{\sigma}\) denotes a Gaussian kernel and \(*\) denotes convolution. Bead-like structures are enhanced using a difference-of-Gaussians response with inner scale \(\sigma_{\mathrm{S}}\) and outer scale \(\sigma_{\mathrm{L}}\), with \(\sigma_{\mathrm{S}}<\sigma_{\mathrm{L}}\),
\begin{equation}
R_k(\mathbf{r}_{\mathrm{o}})=\max\Bigl(0,\,\bigl(G_{\sigma_{\mathrm{S}}}*F_k\bigr)(\mathbf{r}_{\mathrm{o}})-\bigl(G_{\sigma_{\mathrm{L}}}*F_k\bigr)(\mathbf{r}_{\mathrm{o}})\Bigr).
\label{eqAssayDoG}
\end{equation}

Candidate bead centres are detected as regional maxima after suppressing weak maxima and applying a global response threshold. A global threshold is defined as a high percentile of the response distribution
\begin{equation}
t_{\mathrm{g},k}=\mathrm{Percentile}_{p}\bigl(R_k(\mathbf{r}_{\mathrm{o}})\bigr),
\label{eqAssayGlobalThresh}
\end{equation}
where \(p\in(0,100)\). Weak-maxima suppression is implemented by an \(h\)-maxima operator, denoted \(\mathcal{H}(R_k,h_{\max})\), which removes maxima whose height above the local background is below \(h_{\max}\). The candidate mask is then
\begin{equation}
M_k^{\mathrm{cand}}(\mathbf{r}_{\mathrm{o}})=
\mathrm{RegMax}\bigl(\mathcal{H}(R_k,h_{\max})\bigr)(\mathbf{r}_{\mathrm{o}})
\,\mathbb{I}\bigl(R_k(\mathbf{r}_{\mathrm{o}})>t_{\mathrm{g},k}\bigr),
\label{eqAssayCandMask}
\end{equation}
where \(\mathrm{RegMax}(\cdot)\) denotes the regional-maxima operator and \(\mathbb{I}(\cdot)\) is the indicator function. Large artefacts are excluded using an artefact mask computed by thresholding \(F_k\) at a high percentile, followed by basic morphological operations and dilation, and candidates located inside the artefact mask or on the image boundary are rejected.

The remaining candidates are processed in descending order of \(R_k\) and accepted greedily with a minimum separation distance \(d_{\min}\) enforced between accepted candidates. Each candidate at location \(\mathbf{r}_{0}\) is validated using a local window \(\mathcal{W}(\mathbf{r}_{0})\) extracted from \(F_k\). A robust local scale is computed using the median absolute deviation,
\begin{equation}
\mathrm{MAD}\bigl(F_k\mid_{\mathcal{W}}\bigr)=
\mathrm{median}_{\mathbf{r}_{\mathrm{o}}\in\mathcal{W}}
\left|
F_k(\mathbf{r}_{\mathrm{o}})
-
\mathrm{median}_{\mathbf{u}\in\mathcal{W}}F_k(\mathbf{u})
\right|,
\label{eqAssayMAD}
\end{equation}
with corresponding scale estimate
\begin{equation}
\hat{\sigma}_{\mathrm{rob}}=1.4826\,\mathrm{MAD}\bigl(F_k\mid_{\mathcal{W}}\bigr),
\label{eqAssaySigmaRob}
\end{equation}
where the factor \(1.4826\) converts the median absolute deviation to a standard-deviation-equivalent scale under a Gaussian assumption. Two candidate-specific thresholds are then defined. The first threshold uses a local background-plus-margin rule
\begin{equation}
t_{1}=\mathrm{median}\bigl(F_k\mid_{\mathcal{W}}\bigr)+\kappa_{\mathrm{loc}}\hat{\sigma}_{\mathrm{rob}},
\label{eqAssayT1}
\end{equation}
where \(\kappa_{\mathrm{loc}}>0\) controls the margin in units of \(\hat{\sigma}_{\mathrm{rob}}\). The second threshold uses a peak-proportional rule
\begin{equation}
t_{2}=\alpha_{\mathrm{peak}}\,R_k(\mathbf{r}_{0}),
\label{eqAssayT2}
\end{equation}
where \(\alpha_{\mathrm{peak}}\in(0,1)\) is a fixed fraction. The effective local threshold is
\begin{equation}
t_{\mathrm{loc}}=\max(t_{1},t_{2}).
\label{eqAssayTloc}
\end{equation}
A local binary support is then formed as
\begin{equation}
S_k(\mathbf{r}_{\mathrm{o}})=\mathbb{I}\bigl(F_k(\mathbf{r}_{\mathrm{o}})>t_{\mathrm{loc}}\bigr),
\qquad
\mathbf{r}_{\mathrm{o}}\in\mathcal{W}(\mathbf{r}_{0}),
\label{eqAssayLocalSupport}
\end{equation}
and only the connected component of \(S_k\) that contains \(\mathbf{r}_{0}\) is retained by morphological reconstruction. Geometric constraints are applied to the retained component using its area \(A\), major-axis length \(L\), and eccentricity \(e\). A candidate is accepted if
\begin{align}
A_{\min}\le A &\le A_{\max}, \label{eqAssayAreaGate}\\
L &\le L_{\max}, \label{eqAssayMajorGate}\\
e &\le e_{\max}. \label{eqAssayEccGate}
\end{align}
The accepted bead centres for frame \(k\) are denoted \(\mathcal{P}_k=\{(x_n,y_n)\}_{n=1}^{N_k}\), and the frame-level bead count is
\begin{equation}
N_k=\bigl|\mathcal{P}_k\bigr|.
\label{eqAssayCount}
\end{equation}

To summarise temporal dynamics, the normalised fraction of beads remaining on the substrate is defined as
\begin{equation}
f_k=\frac{N_k}{N_0},
\label{eqAssayFrac}
\end{equation}
where \(N_0\) is the bead count in the initial frame. When a smooth monotone kinetic model is required, the temporal evolution is fitted using the Hill-type decay function
\begin{equation}
f(t)=f_{\mathrm{final}}+\frac{f_0-f_{\mathrm{final}}}{1+\left(\frac{t}{t_{1/2}}\right)^n},
\label{eqAssayHill}
\end{equation}
where \(f_0\) and \(f_{\mathrm{final}}\) denote the initial and final bound fractions, \(t_{1/2}\) is the characteristic half-time, and \(n\) is the Hill coefficient that controls the steepness of the transition, which were all treated as fit parameters.

\backmatter

\section*{Funding and acknowledgements}

This research is supported by the Ministry of Education, Singapore, under its Research Centre of Excellence award to the Institute for Digital Molecular Analytics \& Science, NTU (IDMxS, grant: 17435 EDUNC-33-18-279-V12) and Academic Research Fund (Tier 1) Grant No. RG137/24.

\section*{Competing interests}

The authors declare no competing interests.

\section*{Data and code availability}

The authors declare that the data supporting the findings of this study are available within the paper. Any code that supports the findings of this study is available from the corresponding authors upon request.

\end{document}